\documentclass[acus,
              ]{jacow}

\makeatletter
\ifboolexpr{bool{xetex}}                 
 {\renewcommand{\Gin@extensions}{.pdf,%
                    .png,.jpg,.bmp,.pict,.tif,.psd,.mac,.sga,.tga,.gif,%
                    .eps,.ps,%
                    }}{}
\makeatother

\ifboolexpr{bool{xetex} or bool{luatex}} 
 {}                                      
 {\usepackage[utf8]{inputenc}}           

\usepackage[USenglish]{babel}

\ifboolexpr{bool{jacowbiblatex}}
 {%
  \addbibresource{jacow-test.bib}
  \addbibresource{biblatex-examples.bib}
 }{}

\newcommand{\q}[2]{\ensuremath{#1\ \mathrm{#2}}} 
\newcommand{\vol}[1]{\textbf{#1}} 
\newcommand{\be}{\begin{equation}}
\newcommand{\ee}{\end{equation}}
\newcommand{\bea}{\begin{eqnarray}}
\newcommand{\eea}{\end{eqnarray}}
\newcommand{\dt}{\ensuremath{\Delta t}}

\begin{document}

\title{Measurements of beam halo diffusion and \protect\\ population density in
  the Tevatron \protect\\ and in the Large Hadron Collider\thanks{Fermilab is
    operated by Fermi Research Alliance, LLC under Contract
    No.~DE-AC02-07CH11359 with the United States Department of Energy.
    This work was partially supported by the US DOE LHC Accelerator
    Research Program (LARP) and by the European FP7 HiLumi LHC Design
    Study, Grant Agreement 284404. Report number:
    FERMILAB-CONF-14-450-AD-APC.}}

\author{G.~Stancari\thanks{Email:
    $\langle$stancari@fnal.gov$\rangle$.}, Fermilab, Batavia, IL
  60510, USA}

\maketitle

\begin{abstract}
  Halo dynamics influences global accelerator performance: beam
  lifetimes, emittance growth, dynamic aperture, and collimation
  efficiency. Halo monitoring and control are also critical for the
  operation of high-power machines. For instance, in the
  high-luminosity upgrade of the LHC, the energy stored in the beam
  tails may reach several megajoules. Fast losses can result in
  superconducting magnet quenches, magnet damage, or even collimator
  deformation. The need arises to measure the beam halo and to remove
  it at controllable rates. In the Tevatron and in the LHC, halo
  population densities and diffusivities were measured with collimator
  scans by observing the time evolution of losses following small
  inward or outward collimator steps, under different experimental
  conditions: with single beams and in collision, and, in the case of
  the Tevatron, with a hollow electron lens acting on a subset of
  bunches. After the LHC resumes operations, it is planned to compare
  measured diffusivities with the known strength of transverse damper
  excitations. New proposals for nondestructive halo population
  density measurements are also briefly discussed.
\end{abstract}

\section{Introduction}

Understanding particle losses and beam quality degradation is one of
the fundamental aspects in the design and operation of
accelerators. From the point of view of machine protection, losses
must be absorbed by the collimation system to avoid damaging
components. Beam lifetimes and emittance growth determine the
luminosity of colliders. Knowledge of the machine aperture (physical
and dynamical) and of the mechanisms that drive particle loss is
essential.

The LHC and its planned luminosity upgrades (HL-LHC) represent huge
leaps in the stored beam energy of colliders. In 2011, the Tevatron
stored a beam of 2~MJ at 0.98~TeV, whereas the LHC reached 140~MJ in
2012 at 4~TeV. The nominal LHC will operate at 362~MJ at 7~TeV in
2015, and the HL-LHC project foresees that around 2023 the machine
will store proton beams of 692~MJ.

No scrapers exist in the LHC for full beam at top energy. Moreover,
the minimum design HL-LHC lifetimes (about 0.2~h for slow losses
during squeeze and adjust) are close to the plastic
deformation of primary and secondary collimators.

Halo populations in the LHC are not well known. Collimator
scans~\cite{Burkart:PhD:2012, Valentino:PRSTAB:2013}, van-der-Meer
luminosity scans~\cite{CMS:PAS:2011}, and losses during the
ramp~\cite{Redaelli:IPAC:2013} indicate that the tails above 4$\sigma$
(where $\sigma$ is the transverse rms beam size) represent between
0.1\% and 2\% of the total population, which translates to megajoules
of beam at 7~TeV. Quench limits, magnet damage, or even collimator
deformation will be reached with fast
losses~\cite{Schmidt:IPAC:2014}. In HL-LHC, these fast losses include
crab-cavity failures, which generate orbit drifts of about
2$\sigma$~\cite{Yee-Rendon:PRSTAB:2014}.

Hence, the need arises to measure and monitor the beam halo, and to
remove it at controllable rates. For HL-LHC, beam halo monitoring and
control are one of the major risk factors for operation with crab
cavities. Hollow electron lenses were proposed as an established and
flexible tool for controlling the halo of high-power
beams~\cite{Stancari:CDR:2014}.

The dynamics of particles in an accelerator can be quite
complex. Deviation from linear dynamics can be large, especially in
the beam halo. Lattice resonances and nonlinearities, coupling,
intrabeam and beam-gas scattering, and the beam-beam force in
colliders all contribute to the topology of the particles' phase
space, which in general will include regular and chaotic regions, and
resonant islands. In addition, various noise sources are present in a
real machine, such as ground motion (resulting in orbit and tune
jitter) and ripple in the radiofrequency and magnet power supplies. As
a result, the macroscopic motion can acquire a stochastic character,
which can be described in terms of particle
diffusion~\cite{Lichtenberg:1992, Chen:PRL:1992,
  Gerasimov:FERMILAB:1992, Zimmermann:PA:1994, Sen:PRL:1996}.

Calculations of lifetimes, emittance growth rates, and dynamic
aperture from various sources are routinely performed in the design
stage of all major accelerators, providing the foundation for the
choice of operational machine parameters. Experimentally, it was shown
that beam halo diffusion can be measured by observing the time
evolution of particle losses during a collimator
scan~\cite{Seidel:1994}. These phenomena were used to estimate the
diffusion rate in the beam halo in the SPS at
CERN~\cite{Burnod:CERN:1990, Meddahi:PhD:1991}, in HERA at
DESY~\cite{Seidel:1994}, and in RHIC at
BNL~\cite{Fliller:PAC:2003}. An extensive experimental campaign was
carried out at the Tevatron in~2011~\cite{Stancari:IPAC:2011,
  Stancari:HB:2012, Stancari:BB:2013} to
characterize the beam dynamics of colliding beams and to study the
effects of the novel hollow electron beam collimator
concept~\cite{Stancari:PRL:2011}. Following the results of the Tevatron
measurements, similar experiments were done in the
LHC~\cite{Valentino:MDNote:2012, Valentino:PRSTAB:2013}.

In this paper, we review some of the present and future experimental
methods to estimate beam halo populations, with a discussion of their
systematic effects. We also survey the experimental data on the dynamics of
the beam halo, with a discussion on the relationship between
diffusivities and population densities.

\section{Halo population density}

\subsection{Collimator scans}

The dynamics of the beam halo was studied experimentally with
collimator scans~\cite{Seidel:1994} at the Fermilab Tevatron
proton-antiproton collider in~2011. The main motivation was to observe
the effect on diffusion of beam-beam forces and of the novel hollow
electron beam collimator~\cite{Stancari:PRL:2011}. The same data was
used to estimate halo populations beyond about 7$\sigma$. Lower
amplitudes could not be reached because of the minimum size of the
collimator steps and of the safety thresholds of the beam loss
monitors.

In the Tevatron, 36 proton bunches (identified as P1--P36) collided
with 36 antiproton bunches (A1--A36) at the center-of-momentum energy
of 1.96~TeV. There were 2 head-on interaction points (IPs),
corresponding to the CDF and the DZero experiments. Each particle
species was arranged in 3~trains of 12~bunches each, circulating at a
revolution frequency of 47.7~kHz. The bunch spacing within a train was
396~ns, or 21 53-MHz rf buckets. The bunch trains were separated by
2.6-$\mu$s abort gaps. The synchrotron frequency was 34~Hz, or
$7\times 10^{-4}$ times the revolution frequency. The machine operated
with betatron tunes near~20.58.  Protons and antiprotons shared a
common vacuum pipe. Outside of the interaction regions, their orbits
wrapped around each other in a helical arrangement. Therefore, bunch
centroids could be several millimeters away from the physical and
magnetic axes of the machine. Beam intensities and bunch lengths were
measured with a resistive wall monitor. Transverse beam sizes were
inferred from the recorded synchrotron light images.

All collimators were retracted except one. The collimator of interest
was moved in or out in small steps, and the corresponding local loss
rates were recorded as a function of time. A detailed description of
the Tevatron collimation system can be found in
Ref.~\cite{Mokhov:JINST:2011}.

Collimator scans were also used to estimate halo populations in the
LHC at 4~TeV. The experiments were described in
Refs.~\cite{Valentino:MDNote:2012, Valentino:PRSTAB:2013}. The goal of
these experiment was to measure both halo populations and
diffusivities under the same conditions. One nominal bunch
(\q{1.15\times 10^{11}}{protons}) per beam was used. The study started
with squeezed, separated beams. Orbit stabilization was turned
off. The primary and secondary collimators in the IR7 region were
retracted from their nominal settings of 4.3$\sigma$ and 6.3$\sigma$
respectively to a half gap of 7$\sigma$. The jaws of a vertical and a
horizontal primary collimators were moved in small steps. The
collimators were selected from different beams to be able to perform
the scrapings in parallel without inducing cross-talk in the
loss-monitor signals. The jaws were moved after waiting for the beam
losses from the previous step to reach a steady-state (approximately
every 10 to 40 seconds). The jaws were left for a few minutes in the
beam after they had reached their final inward position, to allow the
losses to stabilize. Subsequently, the jaws were moved out in small
steps, again after waiting for the transient to decay. The procedure
of inward and outward steps was then repeated after bringing the beams
into collision.

\begin{table*}[tbh]
\centering
\caption{Estimates of halo population in the LHC at 4~TeV with
  collimator scans.}
\begin{tabular}{ccccccc}
\toprule
Data set  & Beam & Plane & Collisions? & Action~$J$ corresponding & \multicolumn{2}{c}{Tail
population beyond 4$\sigma$} \\
 & & & & to 4$\sigma$ [$\mu$m] & [$10^8\ p$] & [\%] \\
\midrule
1 & B1 & V &  N & 0.00246 & 8.2 & 0.58 \\
2 & B1 & V &  Y & 0.00226 & 1.6 & 0.13 \\
3 & B2 & H &  N & 0.00492 & 8.5 & 0.86 \\
4 & B2 & H &  Y & 0.00347 & 2.4 & 0.35 \\
\bottomrule
  \end{tabular}
\label{tab:halo.pop.table}
\end{table*}

\begin{figure*}
\centering
\includegraphics*[width=\textwidth]{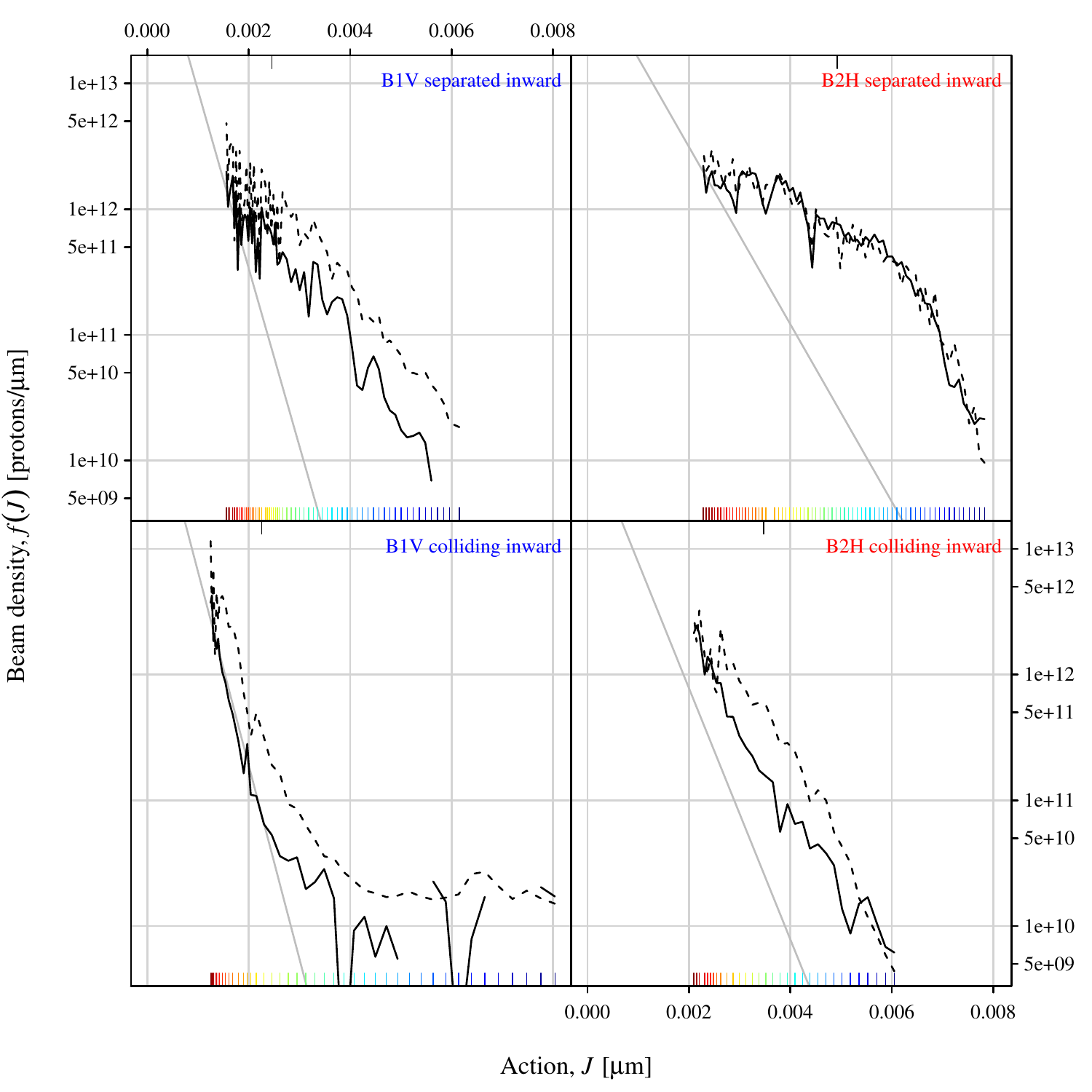}
\caption{Measured beam halo distributions during the inward collimator
  scans in the LHC at 4~TeV, as a function of collimator position in
  units of action~$J \equiv x_c^2 / (2\beta)$, where $x_c$ is the half
  gap and $\beta$ is the local amplitude lattice function: from total
  intensity loss (solid black); from integrated loss-monitor rates
  (dashed black). The gray line denotes a Gaussian core with the
  measured beam emittance. The colored vertical tick marks indicate
  the collimator positions. The 2 left plots refer to the vertical
  scraping of beam~1 with separated and colliding bunches; the 2 right
  plots are for beam~2 in the horizontal plane.}
\label{fig:beam-pop}
\end{figure*}

In the approximation of static beam distributions, the beam densities
can be calculated from the measured intensity loss during a short
interval (4~s in this case) centered around the collimator
movement. The results are shown in Figure~\ref{fig:beam-pop} (solid
black lines). Similar results are obtained by integrating the
calibrated losses over the same short interval
(Figure~\ref{fig:beam-pop}, dashed black lines). For comparison, a
Gaussian core with the measured beam emittance is also shown in
Figure~\ref{fig:beam-pop} (gray line). Tail populations beyond
4$\sigma$ are reported in Table~\ref{tab:halo.pop.table}. It is
interesting to note the depletion in the case of collisions compared
with separated beams.

Scans with primary collimators in a dispersive region of the LHC were
used to estimate the population of the off-momentum halo and of the
abort gap~\cite{Mirarchi:CERN-ACC:2014}. Tails of about 0.5\% were
observed at a relative momentum deviation above $1\times 10^{-3}$.

\subsection{Other estimates}

An estimate of the beam distribution can be obtained with van-der-Meer
luminosity scans~\cite{CMS:PAS:2011, Yee-Rendon:PRSTAB:2014}. The two
colliding beams are displaced with respect to each other, and the
luminosity is recorded as a function of separation. With this
technique, it was observed that the luminosity curve is well described
by a double Gaussian, and that the beam population above 4$\sigma$ was
of the order of 0.1\%.

Losses during the LHC acceleration ramp, as the collimator settings
are tightened, also give indications of the magnitude of beam
tails~\cite{Redaelli:IPAC:2013}. On average, 1\% of the beam was above
6$\sigma$ and was lost during the ramp in~2012.

\subsection{Nondestructive beam halo diagnostics}

Halo monitoring is clearly a high priority for high-power machines. A
true halo monitor should provide a real-time, 2-dimensional transverse
beam distribution. This requires a response time of a few seconds, and
a dynamic range of the order of $10^6$.

In the LHC, it is planned to use synchrotron radiation as diagnostic
phenomenon~\cite{Schmickler:private:2014}. Dynamic range can be
achieved with the coronagraph technique~\cite{Mitsuhashi:DIPAC:2005}
(perhaps replacing the stop with a neutral filter), or more simply
with a set of state-of-the-art digital cameras.

A new kind of detector was recently developed for the RHIC electron
lenses~\cite{Thieberger:IBIC:2014}. It was shown that the rate of
electrons backscattered towards the gun by Coulomb collisions with the
circulating ions is a sensitive probe of the overlap between the two
beams. Although a 2-dimensional reconstruction would require some kind
of scanning of the electron beam, the method is based on scintillator
counters and has a wide dynamic range. It is a promising means to
continuously monitor the halo, especially in conjunction with a hollow
electron lens.

\section{Halo diffusivity}

Halo diffusivities can also be measured with collimator
scans~\cite{Seidel:1994}. All collimators except one are retracted.
As the collimator jaw of interest is moved in small steps (inward or
outward), the local shower rates are recorded as a function of
time. Collimator jaws define the machine aperture. If they are moved
towards the beam center in small steps, typical spikes in the local
shower rate are observed, which approach a new steady-state level with
a characteristic relaxation time. When collimators are retracted, on
the other hand, a dip in loss rates is observed, which also tends to a
new equilibrium level.

We consider the evolution in time~$t$ of a beam of particles with
density~$f(J,t)$ described by the diffusion equation
$\partial_t f = \partial_J \left(D \, \partial_J f \right)$, where~$J$
is the Hamiltonian action and~$D$ the diffusion coefficient in action
space. The particle flux at a given location $J=\bar{J}$ is $\phi = -D
\cdot \left[ \partial_J f \right]_{J=\bar{J}}$.  During a collimator step,
the action~$J_c = x^2_c / (2 \beta_c)$, corresponding to the
collimator half gap~$x_c$ at a ring location where the amplitude
function is~$\beta_c$, changes from its initial value~$J_{ci}$ to its
final value~$J_{cf}$ in a time~\dt. In the Tevatron, typical steps in half gap
were \q{50}{\mu m} in 40~ms; smaller steps (\q{10}{\mu m} in 5~ms,
typically) were possible in the LHC. In both cases, the amplitude
function was of the order of a hundred meters.  It is assumed that the
collimator steps are small enough so that the diffusion coefficient
can be treated as a constant in that region. If~$D$ is constant, the
local diffusion equation becomes $\partial_t f = D \, \partial_{JJ}
f$.  With these definitions, the particle loss rate at the collimator
is equal to the flux at that location: $L = -D \cdot \left[ \partial_J
  f \right]_{J=Jc}$.  Particle showers caused by the loss of beam are
measured with scintillator counters or ionization chambers placed
close to the collimator jaw. The observed shower rate is parameterized
as $S = kL + B$, where~$k$ is a calibration constant including
detector acceptance and efficiency and~$B$ is a background term which
includes, for instance, the effect of residual activation. Under the
hypotheses described above, the diffusion equation can be solved
analytically using the method of Green's functions, subject to the
boundary condition of vanishing density at the collimator and
beyond. Details are given in Ref.~\cite{Stancari:FN:2011}. By using
this diffusion model, the time evolution of losses can be related to
the diffusion rate at the collimator position. With this technique,
the diffusion rate can be measured over a wide range of amplitudes.

Some of the results of measurements in the Tevatron were presented in
Refs.~\cite{Stancari:IPAC:2011, Stancari:HB:2012,
  Stancari:BB:2013}. Experiments in the LHC were reported in
Ref.~\cite{Valentino:PRSTAB:2013}. It was shown that the value of the
diffusion coefficient near the core is compatible with measured
emittance growth rates. The effect of collisions in both the Tevatron
and in the LHC was clearly visible. In the Tevatron, diffusion
enhancement in a specific amplitude region due to a hollow electron
lens was observed. During the next run in 2015, we propose to measure
halo diffusion in the LHC as a function of excitation strength in the
transverse dampers, to provide a further test of the accuracy of the
technique.

\section{Discussion and conclusions}

Extracting beam distributions from collimator scrapings requires some
care. The underlying assumption is that the beam distribution is
static or that, if there is beam diffusion, it is independent of
amplitude. In reality, the diffusion rate increases with betatron
amplitude. Neglecting this fact results in overestimating the tails,
and may explain in part the discrepancy between slow and fast
scrapings in Ref.~\cite{Burkart:PhD:2012}. Another systematic effect
is introduced by using a loss-monitor calibration that is independent
of collimator position.

Van-der-Meer scans are used to measure the effective beam overlap and
the absolute calibration of luminosity. Extracting from them a beam
distribution and a halo population requires further hypotheses. The
assumption that the two beams are identical introduces a systematic
uncertainty.

It may be possible to get more accurate estimates of halo populations
by taking into account the relationship between population density,
diffusivity, and instantaneous loss rates from the diffusion model.

As an example, for simplicity, let us consider the Gaussian core of a
beam with root-mean-square (rms), unnormalized
emittance~$\varepsilon$. In action coordinates, this translates into
an exponential density $f_G(J, t) = (N/\varepsilon)\cdot
\exp{\left[-J/\varepsilon \right]}$.  Let's further assume a constant
intensity decay, $N(t) = N_0 \exp{(-\lambda t)}$, and a constant
emittance growth rate: $\varepsilon = \varepsilon_0 \exp{(\gamma
  t)}$. By multiplying the diffusion equation by~$J$ and integrating,
one obtains a relationship between the emittance growth, the diffusion
coefficient, and its gradient $D' \equiv \partial_J D$: $ \gamma = 2
\left\langle \left(D/\varepsilon - D'\right) \cdot J \right\rangle /
\varepsilon^2$, where $\langle\rangle$ indicates an average over the
distribution function. Moreover, by substituting the Gaussian form of
the density~$f_G$ directly into the diffusion equation, one obtains a
first-order differential equation for the diffusion coefficient:
\be D' - D/\varepsilon + \gamma\cdot J -
(\lambda+\gamma)\varepsilon = 0 \ee
It can be solved by imposing a null flux at the origin. This results
in explicit forms for the diffusion coefficient as a function of
action:
\be D(J) = \gamma\varepsilon J + \lambda\varepsilon^2 \left[
  \exp{\left(J/\varepsilon\right)} - 1
\right].\ee
In other words, an exponentially increasing diffusion coefficient is
necessary to produce a Gaussian beam distribution. In more realistic
cases ($D$ increasing as a power of $J$), beam tails are inevitable.

These relationships can be used to test the stochastic model of halo
dynamics and, if it is verified, to provide more accurate measurements
of halo populations, which take diffusivity into account. One of the
advantages of collimator scans is that they allow a simultaneous
measurement of losses, drift velocities, and diffusivities as a
function of betatron amplitude.

\section{Acknowledgments}

This paper was based on the work of many people. In particular, the
author would like to thank S.~Redaelli, B.~Salvachua, and G.~Valentino
of the LHC Collimation Working Group at CERN and G.~Annala,
T.~R.~Johnson, D.~A.~Still, and A.~Valishev at Fermilab. W.~Fischer,
X.~Gu (BNL), R.~Bruce, H.~Schmickler (CERN), N.~Mokhov, T.~Sen,
V.~Shiltsev (Fermilab), and M.~Seidel (PSI) provided valuable
insights.

%
%





\end{document}